\documentclass{article}

\usepackage{arxiv}

\usepackage[utf8]{inputenc} 
\usepackage[T1]{fontenc}    
\usepackage{url}            
\usepackage{booktabs}       
\usepackage{amsfonts}       
\usepackage{nicefrac}       
\usepackage{microtype}      
\usepackage{lipsum}
\usepackage{graphicx}
\usepackage{tabularx}
\usepackage{booktabs}
\usepackage{tabularx}
\usepackage{multirow}
\usepackage{caption}
\usepackage{graphicx}

\usepackage{subcaption}

\usepackage[sort&compress, numbers]{natbib}
\usepackage{hyperref}
\usepackage{amsmath}

\title{Effects of the Plan Vélo I and II on vehicular flow in Paris - An Empirical Analysis \thanks{Paper submitted for presentation at the 104\textsuperscript{th} Annual Meeting of the Transportation Research Board, Washington D.C., Jan. 2025}}

\author{
 Elena Natterer \thanks{Corresponding author} \\
  Chair of Traffic Engineering and Control \\
  Technical University of Munich, Germany \\
  \texttt{elena.natterer@tum.de} \\
   \And
 Allister Loder \\
  Professorship for Mobility Policy \\ 
  School of Social Sciences and Technology \\ 
  Technical University of Munich, Germany \\
  \texttt{allister.loder@tum.de} \\
    \And
 Klaus Bogenberger \\
  Chair of Traffic Engineering and Control \\
  Technical University of Munich, Germany \\
  \texttt{klaus.bogenberger@tum.de} \\
}

\begin{document}
\maketitle

\begin{abstract}
In recent years, Paris, France, transformed its transportation infrastructure, marked by a notable reallocation of space away from cars to active modes of transportation. 
Key initiatives driving this transformation included Plan Vélo I and II, during which the city created over 1,000 kilometres of new bike paths to encourage cycling. For this, substantial road capacity has been removed from the system. This transformation provides a unique opportunity to investigate the impact of the large-scale network re-configuration on the network-wide traffic flow. Using the Network Fundamental Diagram (NFD) and a re-sampling methodology for its estimation, we investigate with empirical loop detector data from 2010 and 2023 the impact on the network's capacity, critical density, and free-flow speed resulting from these policy interventions. We find that in the urban core with the most policy interventions, per lane capacity decreased by over 50\%, accompanied by a 60\% drop in free-flow speed. Similarly, in the zone with fewer interventions, capacity declined by 34\%, with a 40\% reduction in free-flow speed. While these changes seem substantial, the NFDs show that overall congestion did not increase, indicating a modal shift to other modes of transport and hence presumably more sustainable urban mobility. 

\end{abstract}

\section{INTRODUCTION}

The capital of France, Paris, is frequently said to be a pioneer in the reallocation and reconfiguration of road space to other transport modes. But it is not only its pioneering role, but also the speed and magnitude of this transformation that attracts interest. 
Paris plans to completely eliminate local emissions by 2050, making the city emission-free. They also aim to reduce the carbon footprint by 80\% compared to 2004 levels \cite{paris_carbon_neutral}. 

The transformation of Paris has been driven by several political initiatives introduced since 2015, primarily under the leadership of Mayor Anne Hidalgo (2014 - present). Among these initiatives, Plan Vélo I (2015 - 2020) and Plan Vélo II (2021 - 2026) have been instrumental in advancing Paris towards becoming a ``100\% cyclable'' city \cite{buehler}. Between 2015 and 2020, approximately 1,000 km of bike paths were constructed, with plans to add an additional 180 km by 2026 \cite{plan_velo_1, plan_velo_2}. The second phase of the plan also proposes removing 72\% of car parking spaces \cite{remove_parking_spots}. Other notable efforts include the ``Paris Breathes'' program, which began in 2016 and features temporary street closures on Sundays, and the city's shift towards becoming a ``15-minute city'' starting in 2020. This shift includes both conceptual changes, such as converting school playgrounds into parks after hours, and infrastructural updates, like redesigning public squares such as Place de la Bastille to incorporate trees and bike lanes \cite{paris_15_min_city}. These initiatives aim to enhance bike accessibility and promote sustainable transportation options city-wide, fostering a greener, pedestrian-friendly urban environment that is more conducive to cycling.

As a result of these policy interventions, the bike and car network in Paris has experienced significant transformations. These extensive interventions have affected traffic capacity and car behavior within the network, influencing factors such as speed, route choice, and overall traffic dynamics. Paris' transformation stands as a compelling case study for cities confronting emission reduction goals in the face of the climate crisis. The substantial shifts in its transportation infrastructure provide an invaluable opportunity to empirically study their effects on traffic operations and travel behavior, essentially functioning as a real-world experiment. 

The transformation in Paris is of interest for people in practice as seen above, but also in research. For example, exploring whether an impact of the bicycle network and property values exist -  where findings suggest it does not \cite{Paauwe_2021}. Paris exemplifies how a city can maximize the potential of cycling within a metropolitan environment \cite{cycling_book2021}. 

Additionally, a first analysis explored how network changes for bicycle and car traffic relate to travel demand \cite{TrafficReductionParis}. This present analysis differs from this study in the following ways: Firstly, the data has been updated: we now compare the status quo (up to the end of 2023) to that of 2010. This broader timeframe allows us to view the changes in greater perspective and, by including data up to 2023, helps filter out the effects of the COVID-19 pandemic, as conditions largely returned to normal in 2023. Of course, long-term changes, such as the increased prevalence of home-office work, will also have an impact.
With a refined methodology, employing a re-sampling technique for the Network Fundamental Diagram, we can provide more robust estimations for the NFD despite the inherent variability of real-world data. This approach focuses on two specific zones within the network to investigate differences at a more detailed level.
Additionally, we retrieved official data on the cycle paths in Paris, enabling us to conduct a thorough analysis of the changes in cycling infrastructure.
Lastly, we developed a sophisticated methodology for mapping loop detector data to higher-order roads from OpenStreetMap, which allows us to more accurately infer lane-kilometres for the NFD. This enhanced mapping better reflects the actual network conditions and thus improves the precision of our traffic flow estimations. 

Other than that, the impact of these policy interventions on car traffic at the network level has not been extensively explored in the literature. While other studies have investigated the immediate impacts of bicycle traffic on car traffic in Shanghai \cite{IBikeTrafficMFDShanghai}, there is a lack of research on long-term effects similar to those observed in Paris over a 14-year period. 
Existing research primarily focuses on the impact of cycling infrastructure on cycling patterns and bike traffic. One review highlights the role of public policy in promoting biking, emphasizing the need for diverse interventions like infrastructure enhancement and pro-bicycle initiatives \cite{ReviewInfrastructureToIncreaseBiking}. Another study finds that investments in bicycle highways increase cycling commuters and reduce car dependency \cite{ImprovingCyclabilityInNetherlands}. Additionally, research examines the quality of bicycle infrastructure and its effect on bike lane usage \cite{BikeInfrastructureEncourageCycling}. Comparative analysis between classical and recent studies on urban bicycle infrastructure interventions is also conducted \cite{InterventionsInDutchCities}. Finally, studies explore factors contributing to a city's cyclability and introduce methodologies to evaluate cycling quality based on user preferences \cite{UnderstandingBikeability}.

\begin{figure*}[!b]
    \centering
    \includegraphics[width=16cm]{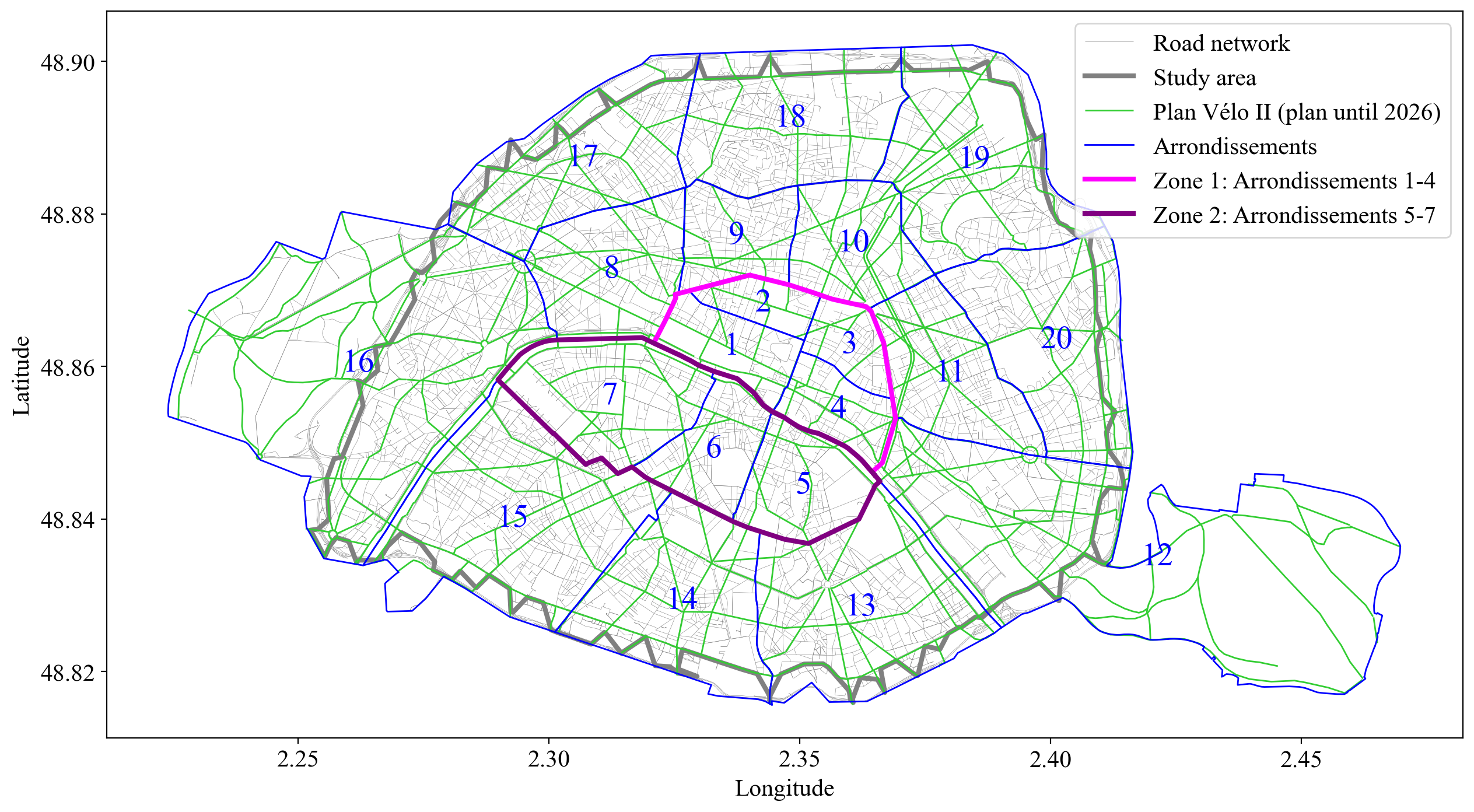}
    \caption{Paris map and its Arrondissements}
    \label{fig:paris_map}
\end{figure*}

The Network Fundamental Diagram (NFD) serves as a novel tool for assessing such policy interventions \cite{Daganzo2007}. The NFD is a reproducible curve between urban traffic density and traffic flow within a network. According to its principles, modifications in the transportation system, particularly in the infrastructure, significantly impact a network’s capacity and critical density. In this paper, we contribute with an empirical application of the NFD to the study of the impact of the policy interventions on network-wide vehicular car traffic in Paris with data from 2010 and 2023. We estimate NFDs using the loops method \cite{Leclercq2014} and the ``re-sampling'' methodology \cite{ambuhl_introducing_2018} to approximate the upper NFD. The re-sampling methodology is applied to homogeneous city zones. We select two zones: The first, Zone 1, characterized by a magnitude of policy interventions (Arrondissements 1 - 4), and the second, Zone 2, where policy interventions occurred at a pace more in line with the rest of the city (Arrondissements 5 - 7). This method yields robust estimations of network capacity and critical density, even under real-world traffic conditions, thereby addressing potential inaccuracies in empirical data. 


Given the considerable number of policy interventions and their implications for the network, it's plausible to expect a decrease in NFD capacity. The change in overall network capacity has implications for car travel times: without change in travel behavior, congestion would worsen substantially, incentivizing car travelers to switch to other modes of transport until a new equilibrium in door-to-door travel times with other modes of transport is reached, following the Downs-Thomson-paradox \cite{mogridge_self-defeating_1997}. Using the NFD as a measure of infrastructure changes, thus, allows us to derive implications for how demand for car trips changes to other modes. Here, recent research has shown how the NFD's critical point is predictable \cite{loder_understanding_2019}, leading to the hypothesis that reducing the overall network length reduces the overall capacity when transforming a car city to a bicycle/public transport city by up to 50\%. Regarding the critical density, the change depends on the presence and mixing with other modes of transport as well as network design: reducing the network car network length, while keeping public transport operations at the same level is expected to increase the critical density as bus operations and priority slow car traffic \cite{Castrillon2018,Geroliminis2014}. Regarding network, reducing the overall route choice or redundancy for car traffic, in other words, forcing car traffic to a limited set of routes, is expected to decrease the critical density. Understanding how the NFD changed then allows us to derive implications for suitable transportation demand management schemes to improve mobility for everyone in the city \cite{balzer_modal_2024,menelaou_convexification_2023}. 

This paper presents a comprehensive large-scale and long-term empirical assessment of network changes and car traffic flow in Paris. Our data-driven analysis reveals that traffic production has decreased by up to 53\%, depending on the specific area of the city examined. Despite this significant reduction in traffic production, congestion levels have remained relatively stable. This suggests that a shift towards more sustainable transport modes, as intended by political initiatives, has indeed occurred.

The remainder of this paper is organized as follows. First, we present the methodology to investigate network and capacity changes using the Network Fundamental Diagram. Next, we present the data from Paris and describe its processing for this analysis. Thereafter we present the results. Ultimately, this paper ends with a discussion of the results as well as the conclusions from the study.

\section{Methodology}
\label{sec:methodology}



We examine the impact of Paris' urban road space transformation on traffic operations using the Network Fundamental Diagram (NFD) as a city-scale assessment method. Table \ref{tab:symbols} lists all symbols used in this analysis.

The NFD provides an aggregated, network-wide perspective on the relationship between the number of vehicles in the network and their average speed and collective production of travel \cite{Geroliminis2008, Daganzo2007}. This relationship arises from a combination of network topology and the dynamics of multimodal traffic operations \cite{Daganzo2008, laval_stochastic_2015, Loder2019SciCap, Geroliminis2014}. Utilising this network-wide perspective facilitates a comprehensive assessment of the impacts of large-scale transportation policies that affect the allocation of road space \cite{ortigosa_analysis_2017,Dantsuji2021,loder_optimal_2022}.

\begin{table}[t!]
    \small
    \centering
    \caption{List of symbols used in this analysis}
    \label{tab:symbols}
    \begin{tabularx}{\textwidth}{ c | c | X  }
    \toprule
    \textbf{Symbol} & \textbf{Unit} & \textbf{Description} \\
    \midrule
    $Z$ & - & Zones considered: $Z = \{Z_1, Z_2\}$ \\
    \midrule
    $Y$ & - & Set of years: $\{2010, 2023\}$\\
    $y$ & - & Year index \\
    $D_y$ & - & Set of days in year $y$ \\
    $d$ & - & Day index \\
    $H$ & - & Set of hours in the day (24-hour clock) with elements $\{5,\dots,22\}$\\
    $h$ & - & Hour index \\
    \midrule
    $N_{y}$ & - & Set of detectors in year $y$ \\
    $N_{yz}$ & - & Set of detectors in year $y$ and zone $z$ \\
    $i$ & - & Detector index \\
    \midrule
    $l_i$ & km & Length of road segment of detectors $i$ \\
    $n_i$ & - & Number of lanes in the road segment of detector $i$ \\
    \midrule
    $o_{ihd}$ & - & Detector occupancy\\
    $q_{ihd}$ & veh/h & Flow of vehicles \\
    $k_{ihd}$ & veh/lane-km & Density \\
    $K_{zhd}$ & veh/lane-km & Aggregated density in zone $z$ \\
    $Q_{zhd}$ & veh/lane-km/h & Aggregated flow in zone $z$ \\
    
    \bottomrule
    \end{tabularx}
\end{table}

To estimate the Network Fundamental Diagram, we use the ``re-sampling'' method \cite{ambuhl_introducing_2018}, which provides robust estimations of network capacity and critical density even under real-world traffic conditions. This method proves particularly useful for assessing alterations using empirical data, as it helps mitigate any potential inaccuracies in the empirical dataset. The idea of the re-sampling approach is to identify the most homogeneous sub-samples of all roads by first creating many random sub-samples of the network, estimating for each an NFD, and extracting the smooth upper bound from the superposition of all NFDs. When the re-sampling parameters are chosen appropriately, all points on the upper bound represent the most homogeneous traffic states. \cite{ambuhl_introducing_2018}
The aim of re-sampling is to approximate the NFD to achieve the best possible performance given the available infrastructure \cite{Ambuhl2021}.

In our analysis, we adopt the NFD's representation of density versus vehicle flow per lane-kilometre. We employ a method that utilises loop detector data and network data from OpenStreetMap (OSM) at different time points. 
Loop detectors, integrated into street infrastructure, provide counts of passing vehicles over time, yielding flow ($q$) in vehicles per hour and vehicle occupancy ($o$) as a percentage. These metrics are recorded across Paris at various measurement locations ($i \in N_y$), every hour ($h \in H$), and every day ($d \in D$). It's important to note that loop detectors (denoted as $N_y$) are typically installed on selected roads, assumed to be representative of the broader network. To ascertain the number of lanes on streets where detectors ($i \in N_y$) are placed, we reconcile traffic data with OSM's network information.

\subsection{Normalization of measurements}
For every year $y \in Y$, consider the set $B$ of those observations which have measured the flow $q$ and the occupancy $o$ at every relevant hour $h \in H$: 
\begin{flalign}
B = (q_{ihd}, o_{ihd}) \quad i \in N_y, h \in H, d \in D_y
\end{flalign}

The measurements are available per link and can result from the aggregation of several detectors on this link, that is, not per lane. Analysis using street images from Paris suggests that each lane of a link is monitored by a detector, that is, if a link has three lanes, one detector is placed on each lane. This results in the following units for $q$ and $o$. Flow $q$ is reported as the number of vehicles per hour and link, that is, unit veh/h, and occupancy $o$ is reported as the average detector occupancy of all detectors on that link during an hour. Note that this measurement is without a unit.

The available Parisian data does inform about the length $l_i$ of the road segment $i$ that is monitored by a set of detectors, but not about the number of lanes $n_i$ at the present location. This information is important for computing the total travel production and speed on that link and capture the network impact of lane removal, for example, if one out of three lanes is reallocated to other modes of transport. In Section \ref{sec:mapping_loop_detectors}, we elaborate on retrieving information for the lane-kilometres from OSM. 

For every detector $i \in N_y$, we are given its length $l_i$ and the number of lanes $n_i$.
For every element in $B$, we normalize $q$ and $o$ in order to get the values per lane-km.
We use the formula for loop methods \cite{Leclercq2014} for the NFD estimation. $Q$ is in units flow per hour and lane-kilometre, aggregated over all detectors in the zone: 
\begin{flalign}
Q_{zhd} = \dfrac{\sum_i l_i \cdot q_{ihd} / n_i }{\sum_i l_i}
\end{flalign}

\begin{flalign}
    k = \dfrac{o}{s}
    \label{eq:occtrans}
\end{flalign}

We convert occupancy, denoted as $o$, into density, represented by $k$, Utilising Equation \ref{eq:occtrans}, along with a scalar $s$. We elaborate on the calibration of this scalar in Section \ref{sec: conversion}. The NFD density $K$ aggregated over zone $z$ can then be computed in units vehicles per lane-kilometre:

\begin{flalign}
\label{eq: k}
K_{zhd} = \dfrac{\sum_i l_i \cdot k_{ihd} }{\sum_i l_i}
\end{flalign}

\subsection{Conversion of occupancy to density}
\label{sec: conversion}
As previously mentioned, detectors in Paris measure flow $q$ per link and average occupancy $o$, where both measures are available only per hour. To derive meaningful implications, occupancy must be transformed into average vehicle density $k$ with units of vehicle per lane-kilometre. 
This is conventionally achieved by a linear transformation using the space-effective vehicle length $s$, which includes the detector and vehicle length. For both, no official values are available and hence we make the assumption that $s=0.0055$ km.

We corroborate this transformation using public data from \href{https://www.tomtom.com/traffic-index/paris-traffic/}{TomTom} on speeds in Paris. The source reports an average speed of approx. 38 km/h at 5 am and approx. 34 km/h at 6 am, that is, before the onset of the morning peak hour. It is not reported whether these values are time-mean or space-mean speeds, presumably time-mean speeds \cite{Ambuhl2017}. Note that the time-mean speed is usually larger than the space-mean speed. 

It's anticipated that in 2023, average speeds closely match those reported by TomTom, potentially even slightly surpassing them, given the likelihood of additional speed reductions. Upon comparing TomTom speeds with calibrated speeds during the same time periods using the NFD's $k$ and $q$ parameters, remarkable proximity is observed: 38.51 km/h at 5 am and 34.17 km/h at 6 am. Hence, $s = 0.0055$ km is appropriate for estimating the density in the Paris region. This value is very close to the scaling factor for an experiment in Yokohama (Japan) by Geroliminis et al. \cite{Geroliminis2008}.

\section{DATA}
\label{sec:data}

This section presents the available data set from Paris. There is an extensive network of loop detectors installed, which are mainly installed for traffic control and congestion identification purposes. They measure traffic flow (number of vehicles passing the detector) and occupancy (fraction of time the detector is occupied by a vehicle) in one-hour intervals since 2010. We only considered data from ``complete days'': Those values for which detectors delivered data for flow and occupancy from 5 am to 11 pm on a weekday. 

In order to infer the number of lanes of the street that a detector measures, we geolocated all spatial information of loop detector positions in reference to the whole road network, as described in Section \ref{sec:mapping_loop_detectors}.

\subsection{Selecting the areas for investigation}

As the ``re-sampling'' methodology works best on homogeneous areas of approximately $5 - 10$ km$^2$, we select areas on which we apply the method. Paris is divided into multiple districts known as ``Arrondissements'' (districts), which we use as a boundary. We employ the method on two zones, depicted in Figure \ref{fig:paris_map}. 

Zone 1 encompasses Arrondissements 1 - 4. These central districts, situated north of the Seine, have been focal points of Parisian policy interventions. They were also the first ones in which measures such as car-free Sundays have been implemented. This zone represents a ``progressive transformation'' of the city.
Zone 2 comprises Arrondissements 5 - 7. Located south of the Seine, these districts have also undergone significant revitalization endeavors. Zone 2 embodies transformation endeavors more in line with the rest of the city.

We investigate the impact of implementing varying degrees of progressive political measures in Paris on traffic and travel behaviour, utilising these two zones as our study areas.

\subsection{Data for lane-kilometres from OpenStreetMap}
\label{sec:mapping_loop_detectors}

To estimate the NFD for lane-kilometres, we need to know the number of lanes on the roads where the detectors are installed. For this purpose, we use data from OpenStreetMap (OSM) for the Parisian road network, specifically from January 1, 2013, and January 1, 2024. Unfortunately, data from earlier periods is not available. Therefore, we utilise lane information from 2013 to infer values for 2010, which sufficiently serves our analytical objectives. 

Loop detectors are typically installed on roads which have the highway classification primary, secondary, tertiary, trunk, or motorway in OSM. Therefore, we focus on the network of those roads, the higher-order road network. 

However, a one-to-one mapping does not exist between the higher-order road network and the detector network. This is due to inaccuracies in OSM but also to different representations of the network - for example, the network in OSM consists of smaller road segments. Therefore, our methodology involves a mapping process based on centroid distances and angles. Specifically, for each detector, we identify the most compatible road segment within the OpenStreetMap (OSM) network in terms of both centroid proximity and angle alignment.

\begin{figure}
    \centering
    \begin{subfigure}[b]{0.48\textwidth}
        \centering
        \includegraphics[width=1.0\textwidth]{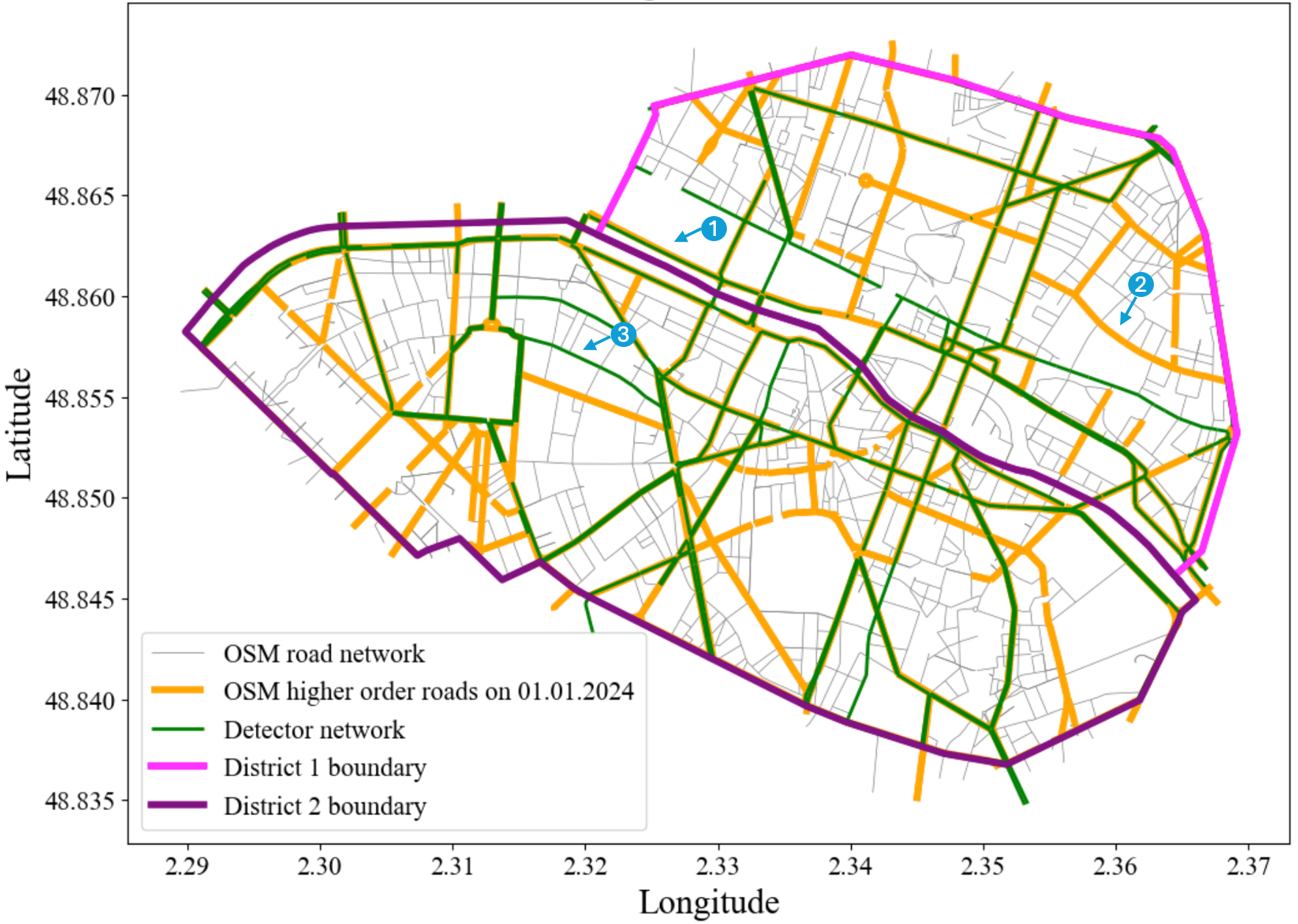}
    \end{subfigure}
    \hfill
    \begin{subfigure}[b]{0.48\textwidth}
        \centering
        \includegraphics[width=1.0\textwidth]{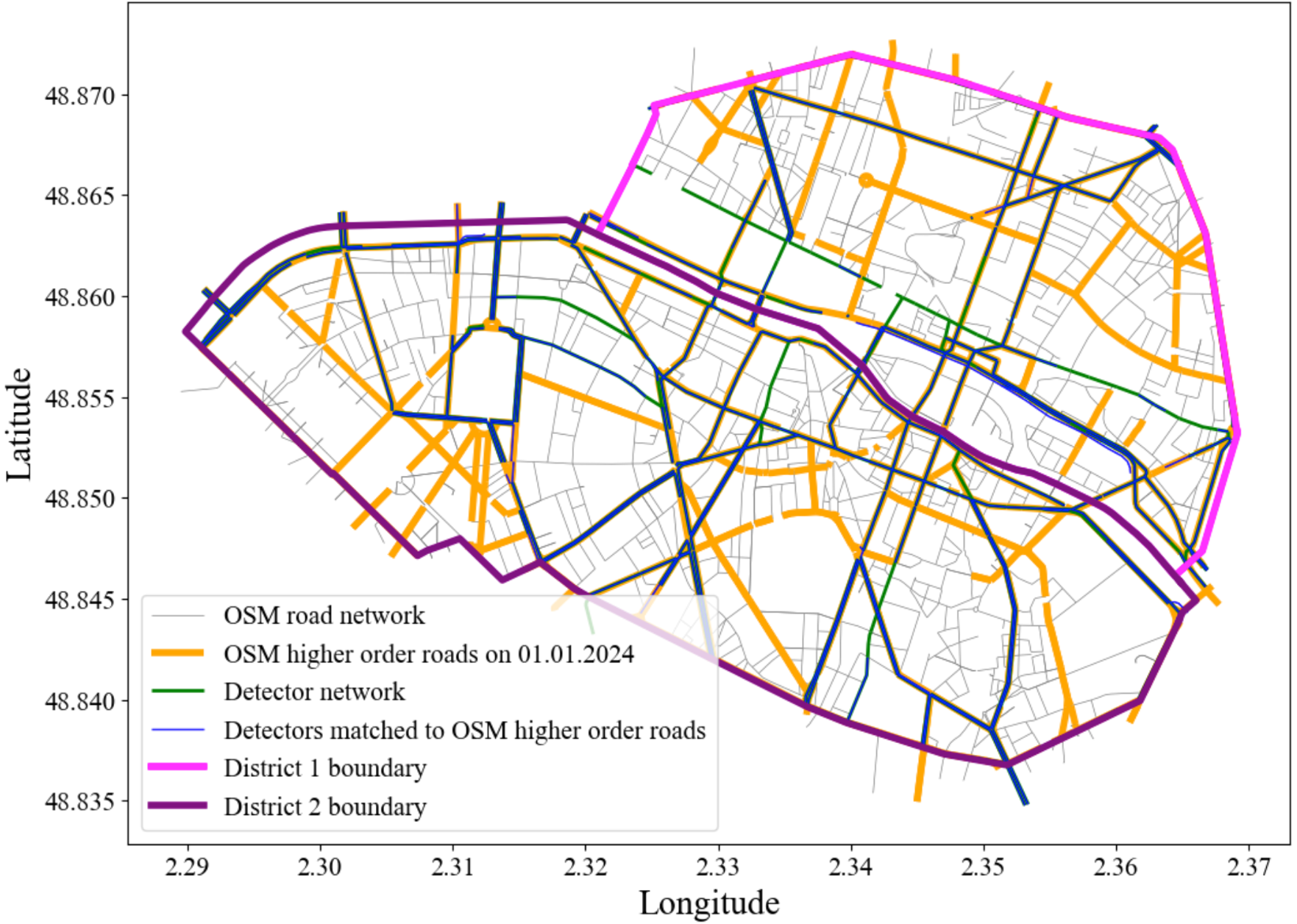}
    \end{subfigure}
    \caption{Matching loop detectors to the Higher-Order Road Network of OpenStreetMap}
    \label{fig:matched_this}
\end{figure} 

In Figure \ref{fig:matched_this} we demonstrate the Higher-Order Road Network from OpenStreetMap and the detector network for the investigated zones 1 and 2. We distinguish the following cases, as indicated by the blue arrows:
\begin{enumerate}
    \item Roads where a detector is located on a road that is also recorded in OpenStreetMap
    \item Higher-Order Roads in OpenStreetMap for which there are no detectors
    \item Detectors where there is no road in OpenStreetMap higher order road: This means that for these roads, we cannot know the number of lanes, so we must approximate them.
    \\
\end{enumerate}
In the matching process, we proceed as follows: First, we retrieve historical data from OpenStreetMap spanning from 2013 - 2024. Secondly, given the dynamic nature of street networks and loop detectors, we repeat the following steps for the years 2013 and 2024. 
\begin{enumerate}
    \item Narrow down the OSM road network to encompass higher-order streets, such as primary, secondary, tertiary, trunk, or motorway. Realistically, loop detectors are seldom installed on other road types (as of January 1st, these were 6352 out of 16554 edges).
    \item Map the detectors to corresponding road segments in OSM based on centroid proximity and angle alignment. 
    \item Define the required accuracy of matches. Notably, not every detector yields a satisfactory mapping.
    \item Utilise the mapped street segments to determine the associated ``highway'' classification and lane count. In cases where lane information is unavailable for certain street segments, we approximate it by using the average of the lane counts per highway type for the given year. For instance, on 1.1.2024, the average lane count on primary streets was 3.09, on secondary streets 2.48, on tertiary streets 2.08, and on trunk roads 4 lanes. \\
\end{enumerate}
Through this process, we estimate the number of lanes for each road segment with detector data. This allows us to accurately approximate the flow and occupancy per lane-kilometre. However, as of January 1,  2024, 331 (12\%) detectors lacked lane specifications, and approximated values were employed. We provide the code for mapping road segments to (historical) data from OpenStreetMap on  
\href{https://github.com/enatterer/Traffic-and-Network-Data-Analysis-Paris}{Github}.



\begin{figure}[htbp]
    \centering
    \begin{minipage}{0.45\textwidth}
        \centering
        \includegraphics[width=\linewidth]{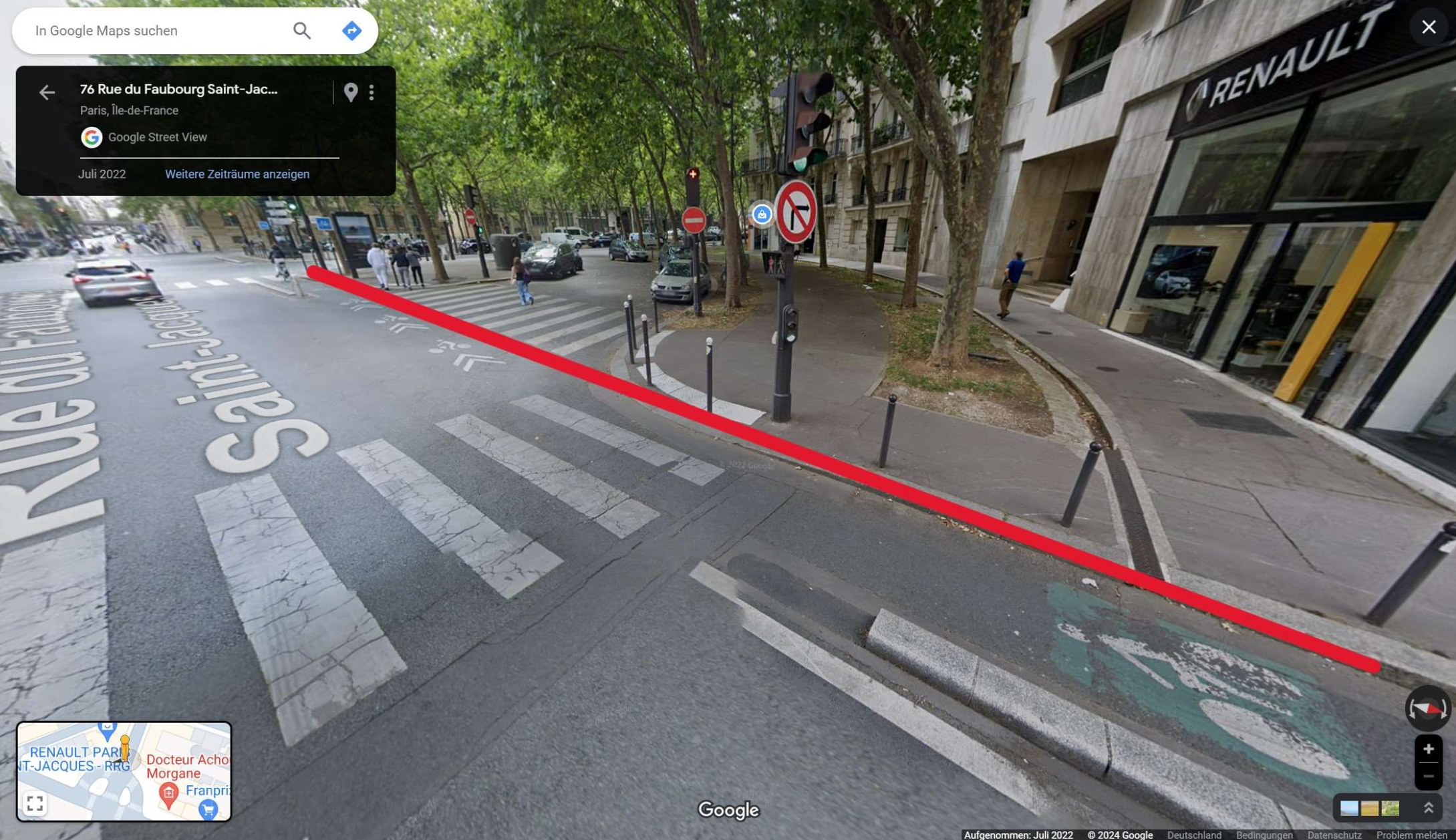}
        \subcaption{``Pistes cyclable''}
        \label{fig:subfig1}
    \end{minipage}\hfill
    \begin{minipage}{0.45\textwidth}
        \centering
        \includegraphics[width=\linewidth]{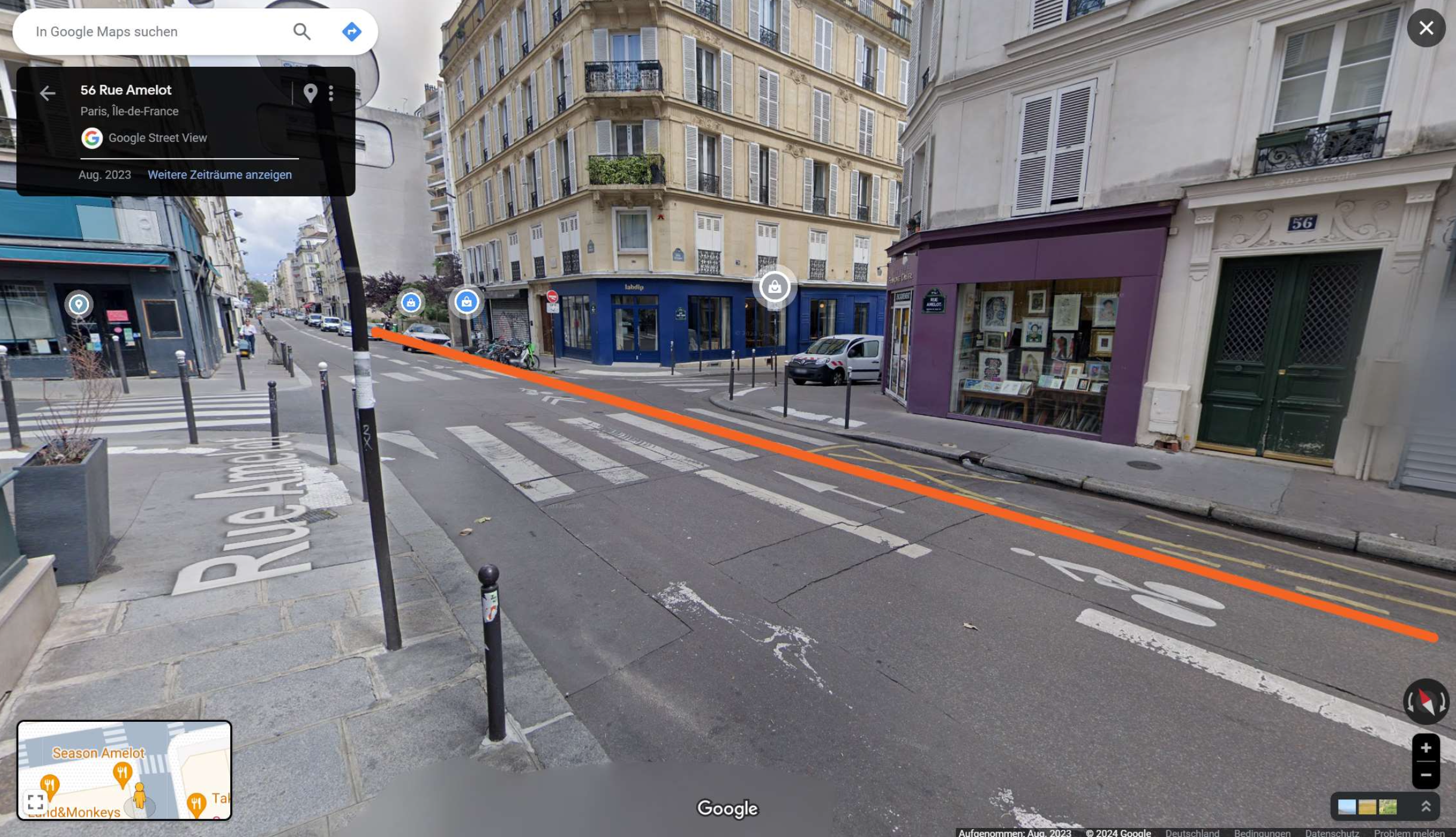}
        \subcaption{``Bande cyclable''}
        \label{fig:subfig2}
    \end{minipage}
    \vspace{0.5cm}
    \begin{minipage}{0.45\textwidth}
        \centering
        \includegraphics[width=\linewidth]{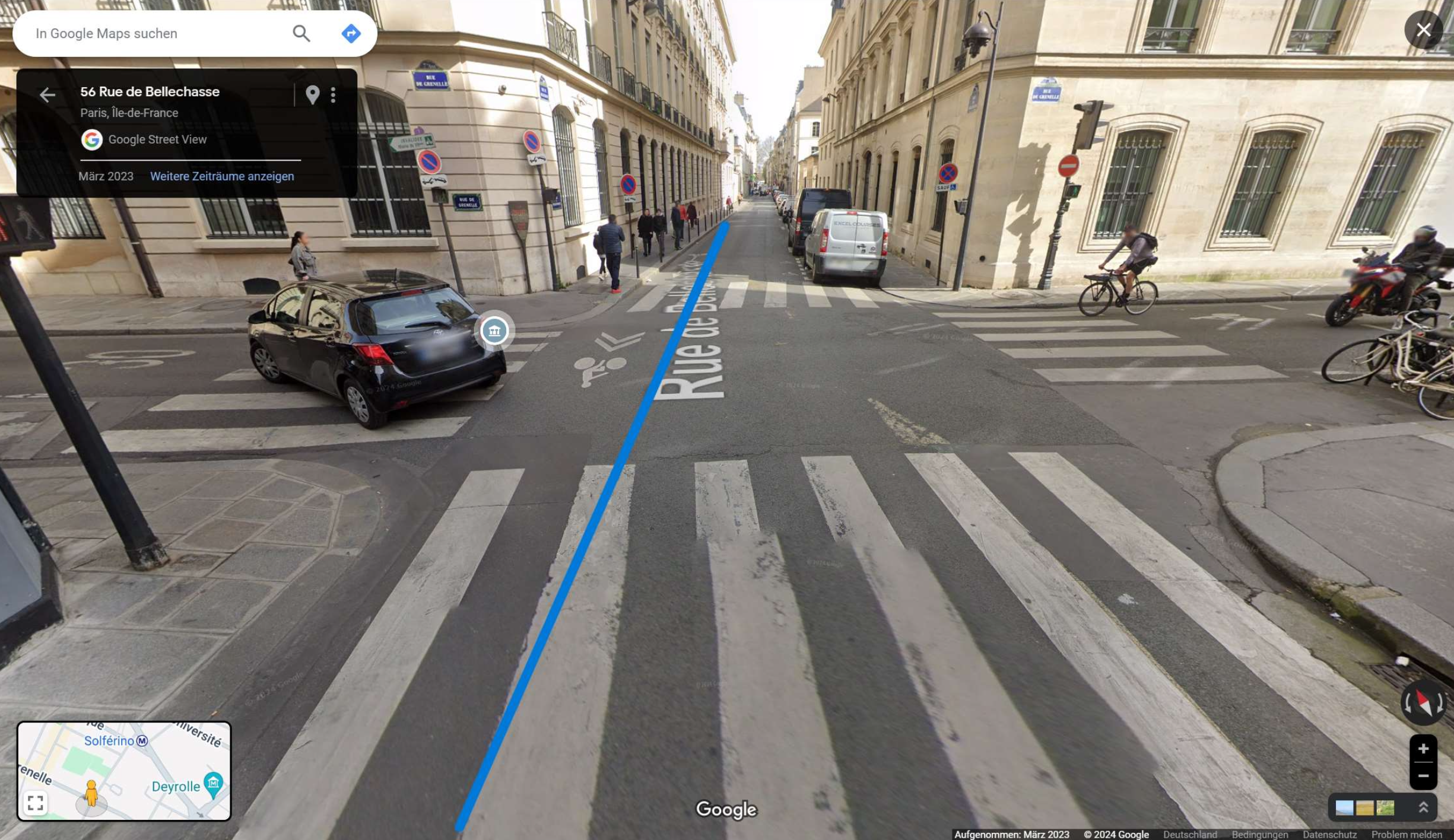}
        \subcaption{``Autre itinéraires cyclables''}
        \label{fig:subfig3}
    \end{minipage}\hfill
    \begin{minipage}{0.45\textwidth}
        \centering
        \includegraphics[width=\linewidth]{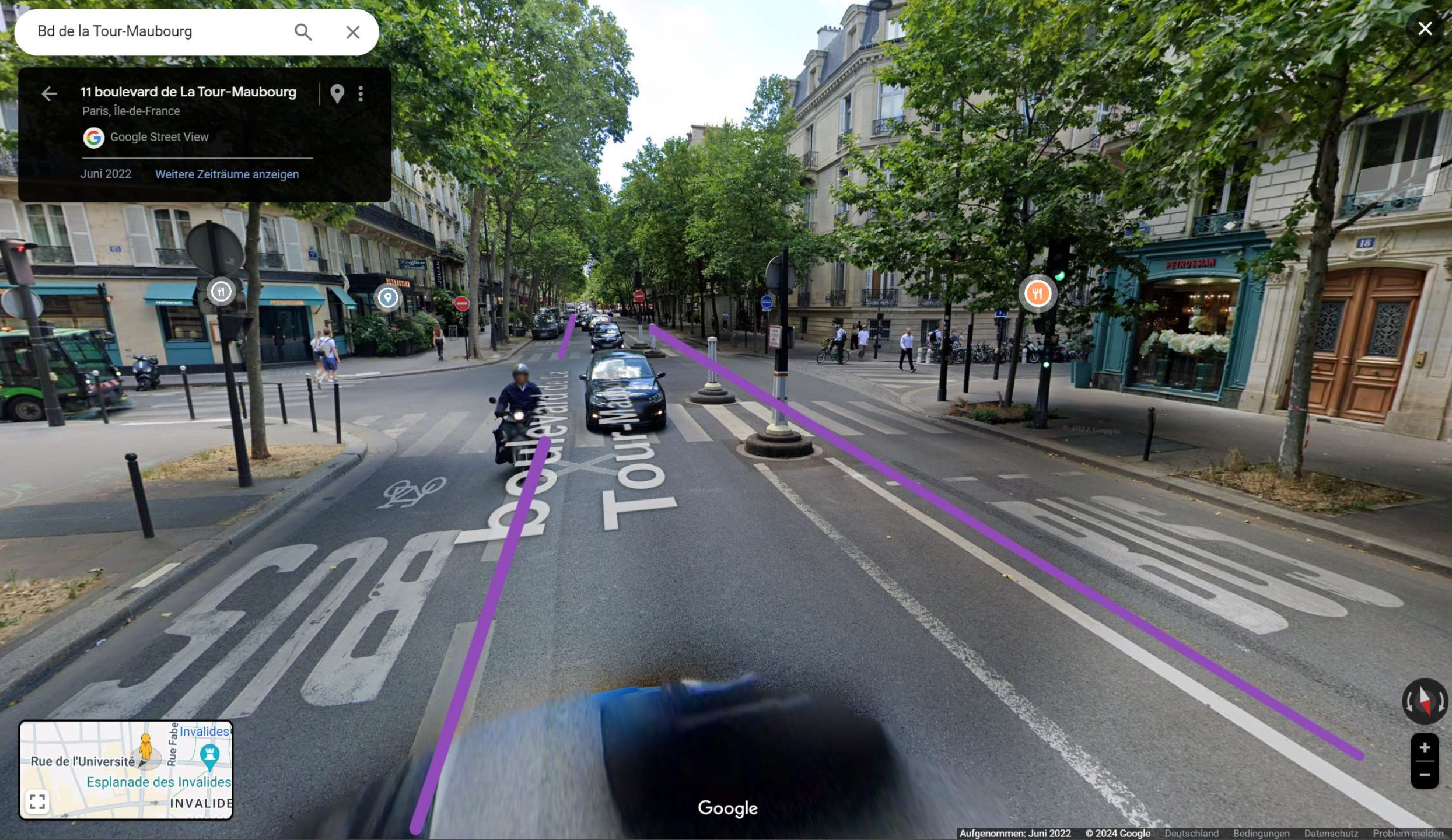}
        \subcaption{``Couloir de bus ouvert aux vélos''}
        \label{fig:subfig4}
    \end{minipage}
    \caption{Example of different types of cycle paths implemented in Paris \protect\footnotemark}
    \label{fig:quadratic_subfigures}
\end{figure}

\subsection{Examining the evolution of the cycling infrastructure}
\label{sec:development_of_bike_network}

We conducted an analysis of the evolution of Paris's bicycle network, using publicly available data \cite{ParisOpenData_Cycleways} spanning from 2000 to 2022. The analysis was conducted using data accessed in April 2024. Regrettably, as of July 2024, this data appears to be unavailable.

\begin{table}[!tb]
    \centering
    \caption{Cycle path classification \cite{kara2024}}
    \label{tab:cyclepath_overview}
    \begin{tabularx}{\textwidth}{ p{4.8cm} | p{3cm} | p{3cm} | X }
    \toprule
    \textbf{Cycle path}  & \textbf{\# of bike lanes} & \textbf{Car lane removal?} & \textbf{Zones} \\
    \midrule
     ``Pistes cyclables'' & 2: bi-directional & Yes, at least one & 30/50 km/h \\
     \midrule
      ``Bande cyclables'' & 1-2  & Sometimes & 30/50 km/h \\
     \midrule
      ``Autre itinéraires cyclables'' &  1-2  & No  & 30 km/h, pedestrian zones\\
      \midrule
     ``Couloir de bus ouvert aux vélos'' & 1: bus lane accessible& No & 30/50 km/h \\
        \bottomrule
    \end{tabularx}
\end{table}

Cycle paths in this official plan are categorized into four types: ``Pistes cyclables'' for lanes exclusively for cyclists with physical separation from other traffic. ``Bande cyclables'' are lanes on roads open to general traffic. ``Couloir de bus ouvert aux vélos'' are bus lanes open to buses and certain categories of vehicles; cyclists are allowed to use it, indicated by bicycle logos on the ground or signs. ``Autre itinéraires cyclables'' encompass all routes that can be used by cyclists, that are closed to general traffic and not part of the other types of described facilities. This includes pedestrian areas, contraflow cycle lanes, and routes closed to general traffic. We provide details on the classifications in table \ref{tab:cyclepath_overview}.

\footnotetext{All figures are taken  from Google Street View. a) is at ``Place Saint Jacques'', b) at ``Rue Amelot'', c) at ``Rue de Bellechasse'', and d) is at ``Boulevard de la Tour Maubourg''.}

Of particular significance for our analysis are the ``Pistes cyclables'', which stand out as the only bike paths featuring physical separation. The creation of these paths consistently involved the removal of at least one car lane.

The development of the bike network length over time is illustrated in Table \ref{tab:bike_network_length}. Plan Vélo II \cite{PlanVeloII_ParisOpenData} is depicted in Figure \ref{fig:paris_map}, illustrating the expansion of the ``Pistes cyclables'' network. This network is projected to span 447 km upon its completion by 2026, in alignment with official reports indicating a coverage of 334 km as of 2022.

\begin{table}[!tb]
    \small
    \centering
    \caption{Bike network length 2010 and 2022 \cite{ParisOpenData_Cycleways}.}
    \begin{tabularx}{12cm}{p{4cm} |c|c|c}
    \toprule
    \textbf{Area \protect\footnotemark} & \textbf{Length 2010} & \textbf{Length 2022} & \textbf{Increase} \\
    \midrule
    Paris region & 422.25 km & 1083.09 km & 257 \% \\
    \midrule
    Zone 1 & 44.37 km & 115.12 km & 259 \%  \\
    \midrule
    Zone 2 & 30.23 km  &  104.46 km & 346 \% \\
    \midrule
    Paris region: ``Pistes cyclables'' & 170.42 km & 333.99 km & 96 \% \\
    \midrule
    Zone 1: ``Pistes cyclables'' & 4.10 km & 19.51 km & 476 \% \\
    \midrule
    Zone 2: ``Pistes cyclables'' & 1.78 km & 16.22 km &  911 \% \\
    \bottomrule
    \multicolumn{4}{l}{\small{``Pistes cyclables'' are physically separated bike paths}}\\
    \end{tabularx}
    \label{tab:bike_network_length}
\end{table}





\section{RESULTS}
\label{sec:results}

\begin{figure*}[!tb]
    \centering
    \includegraphics[width=15cm]{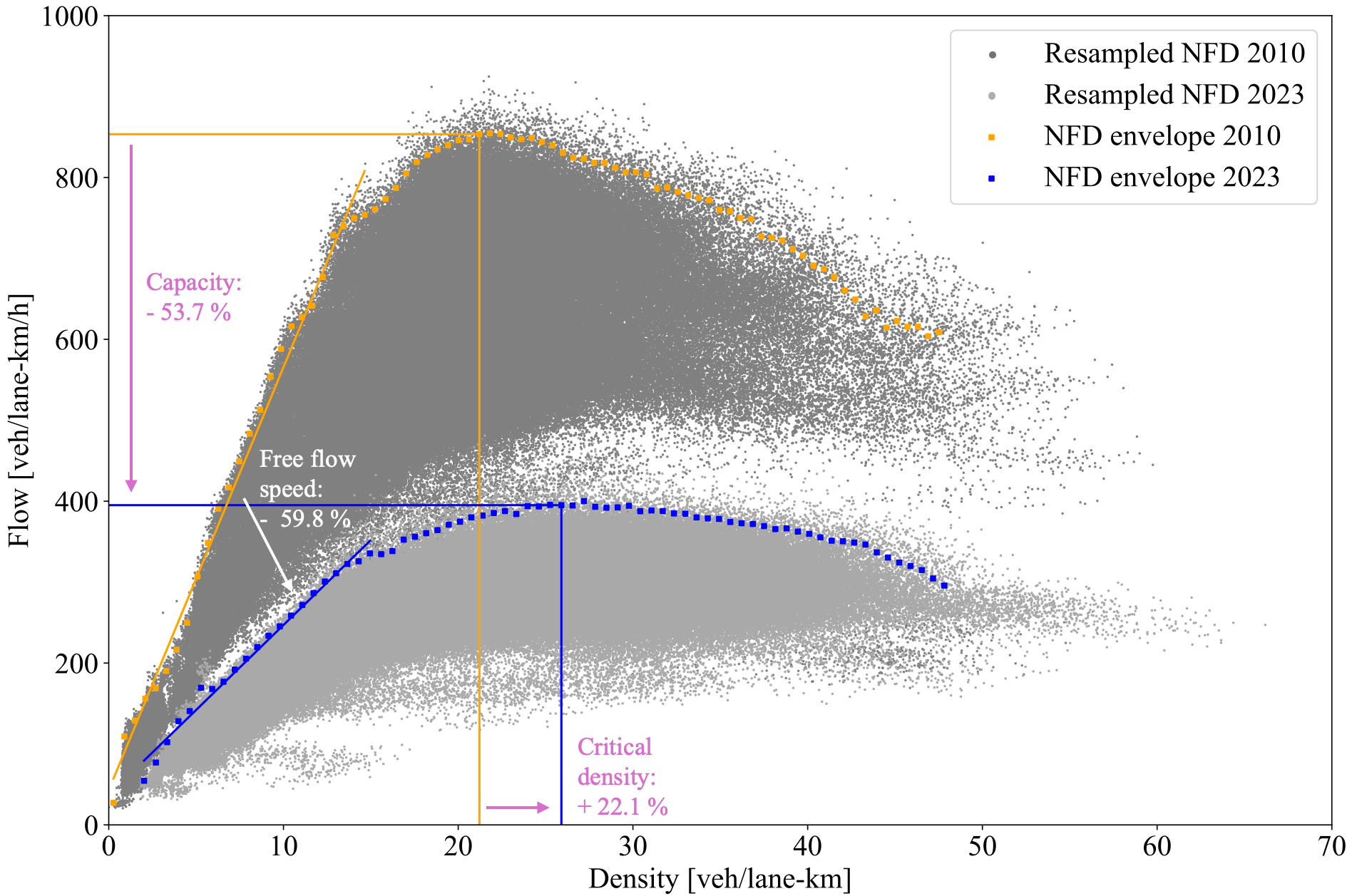}
    \caption{Zone 1 - Resampled NFD 2010 and 2023}
    \label{fig:resampled_NFD_zone_1}
\end{figure*}

\begin{figure*}[!tb]
    \centering
    \includegraphics[width=15cm]{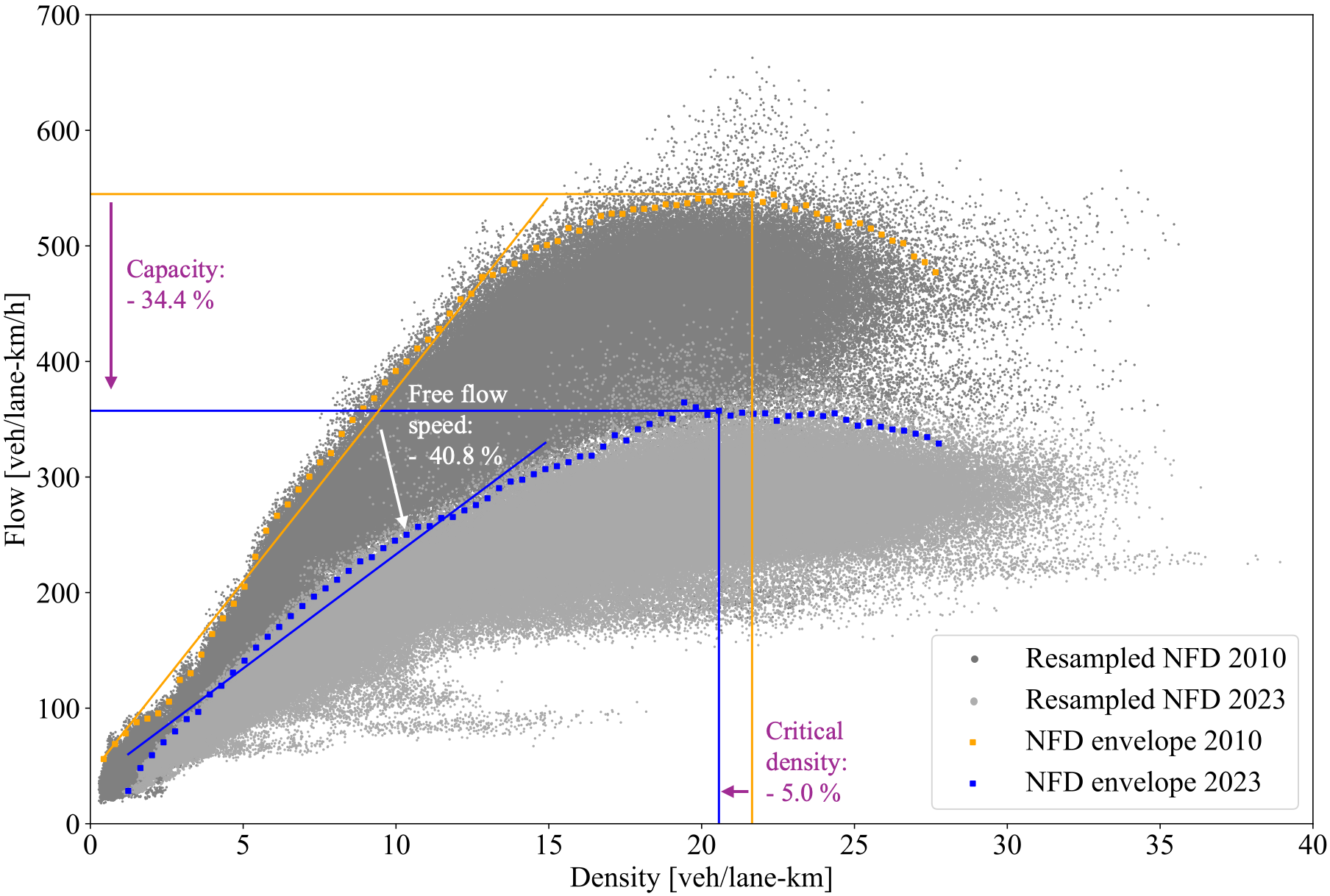}
    \caption{Zone 2 - Resampled NFD 2010 and 2023}
    \label{fig:resampled_NFD_zone_2}
\end{figure*}

In this section, we present the results of applying our methodology from Section \ref{sec:methodology} to the data in Section \ref{sec:data}. In Figure \ref{fig:resampled_NFD_zone_1} and \ref{fig:resampled_NFD_zone_2}, we show the resampled NFD for zones 1 and 2, respectively. It's important to note that every point in the NFD represents a macroscopic traffic state in terms of vehicle density and vehicle flow. We computed the free-flow speed as the speed at a density of 0 - 15 veh/km. The NFD envelope is defined as the median of the top $M$ flow values per density bin, where the value of $M$ depends on the number of observations within each bin. We select $M = 100$ to ensure a smooth upper bound. Using this upper bound, we define capacity as the flow value at the 97.5th percentile to mitigate the influence of outliers. The critical density is the mean density corresponding to this capacity.   \\

Remarkably, both NFDs exhibit a smooth upper bound, which supports the general theory of the NFD, defined as a tight upper bound relatively independent of demand. We present the results in Table \ref{tab:results}.
\begin{table}[!tb]
    \centering
    \caption{Development of NFD Parameters from 2010 to 2023} 
    \begin{tabularx}{\textwidth}{p{1.5cm} | p{4.5cm} | >{\raggedleft\arraybackslash}p{2.5cm} | >{\raggedleft\arraybackslash}p{2.5cm} | >{\raggedleft\arraybackslash}X}
    \toprule
    \textbf{Zone} & \textbf{Metric} & \textbf{Absolute (2010)} & \textbf{Absolute (2023)} & \textbf{Percentual Change} \\ 
    \toprule
    \multirow{3}{*}{Zone 1} & Capacity (veh/lane-km/h) & 854.0 & 395.0 & - 53.7 \% \\
    \cmidrule{2-5}
    & Free-flow Speed (km/h) & 52.3 & 21.0 & - 59.8 \% \\
    \cmidrule{2-5}
    & Critical Density (veh/lane-km) & 21.2 & 26.1 & + 22.1 \% \\
    \midrule
    \multirow{3}{*}{Zone 2} & Capacity (veh/lane-km/h) & 545.0 & 357.0 & - 34.4 \% \\
    \cmidrule{2-5}
    & Free-flow Speed (km/h) & 33.4 & 19.8 & - 40.8 \% \\
    \cmidrule{2-5}
    & Critical Density (veh/lane-km) & 21.7 & 20.6 & - 5.0 \% \\
    \bottomrule
    \end{tabularx}
    \label{tab:results}
\end{table}
For Zone 1, we found that the capacity dropped in the considered time frame from 854 veh/lane-km/h to 395 veh/lane-km/h, a substantial decrease of 53.71\%. The critical density increased from 21 veh/lane-km to 26 veh/lane-km, representing a 22.12\% increase. This phenomenon can be attributed to the fact that the same level of public transport operations is made on less overall network infrastructure, i.e., its relative shares of operations is increasing \cite{Geroliminis2014}. In Zone 2, capacity dropped from 545 veh/lane-km/h in 2010 to 357 veh/lane-km/h in 2023, marking a drop of 34.41\%. The critical density decreased only slightly, from 21.65 veh/lane-km in 2010 to 20.57 veh/lane-km in 2023. This trend is likely attributable to reduced space for cars and an increase in viable alternatives \cite{Castrillon2018}. Both zones experienced a significant drop in free-flow speed. In Zone 1, it dropped from 52.25 km/h in 2010 to 21.00 km/h in 2023, representing a reduction of about 60\%. In Zone 2, it decreased from 33.36 km/h to 19.75 km/h, a decrease of about 40\%. These findings are consistent with our initial hypotheses - both zones experienced a substantial decrease in traffic flow and speed, with Zone 1 seeing particularly significant declines. 



In terms of network supply, the interventions in Paris, including Plan Vélo I and II, led to an expansion of bike lanes and a reduction in space allocated to cars. The cycling network has notably grown, according to cycling paths data from official sources \cite{ParisOpenData_Cycleways}, with detailed findings provided in Table \ref{tab:bike_network_length}. In particular, the establishment of separated bike paths, referred to as ``Pistes cyclables'' in Section \ref{sec:development_of_bike_network}, has notably increased. When analysing the relative increases, it seems that Zone 2 has experienced more significant developments. However, one must not overlook the absolute numbers: Despite its smaller geographical size compared to Zone 2, Zone 1 already had a more extensive bicycle network in 2010. This network comprised both overall routes and distinct cycling paths, which are especially noteworthy.
This suggests that Zone 1 was already more bicycle-friendly and less reliant on cars compared to Zone 2 by 2010. Subsequent network modifications have likely only propelled this progress further.

Regarding the road network, it's reasonable to expect a slight decrease in lane-kilometres. Natterer et al. showcased a 4.9\% reduction in lane kilometres from 2015 to 2023 \cite{TrafficReductionParis}. Pinpointing the exact reduction from 2010 is challenging due to the continuous enhancements in OSM's accuracy, notably considering that data is only available from 2013 onwards.


The code for creating the re-sampling NFDs can be found on \href{https://github.com/enatterer/Traffic-and-Network-Data-Analysis-Paris}{Github}.

\section{DISCUSSION}


The presented results on the changes in Network Fundamental Diagram (NFD) offer an aggregated perspective on network operations. However, there are several key points to consider:

Firstly, these changes were observed in the loop detector data. The impact of network changes on overall car travel, particularly network exit flows, also depends on average trip length. If car travelers with previously short trips switch modes, this reduces maximum trip production and increases average trip length, thereby reducing network exit flows. Conversely, if car travelers with previously long trips, such as inbound commuters, change modes, a reduced maximum trip production and decreased average trip length can increase network exit flows. Essentially, this results in more trips being produced but less car travel overall. Unfortunately, data on the average trip length of car users between 2010 and 2023 is unavailable, preventing definitive conclusions.


Secondly, when it comes to data, using loop detector data may introduce errors. Increasing the coverage of loop detectors and refining the NFD estimation process could mitigate the potential for bias in loop detector selection. Moreover, obtaining accurate information regarding the number of lanes at measurement locations would enhance the accuracy of travel production estimates. This, coupled with improved speed data for NFD calibration, would facilitate investigations into the consistency of occupancy scale interpretation over time. Although OpenStreetMap offers valuable insights, its lack of precise road network details, delays in updates, and limited availability of data starting only from 2013 hinder its verification with official sources. 

Lastly, a comprehensive assessment of Plan Vélo I and II requires consideration of various dimensions. For instance, it is essential to evaluate not only the immediate impacts on traffic but also the broader implications such as potential health benefits from reduced car usage and air pollution, as well as the promotion of alternative modes of transportation like cycling and the potential mitigation of the urban heat island effect through increased green spaces. 
The results cannot be viewed in isolation; during 2010-2023, various events, including the impact of COVID-19 and a significant trend towards remote work, also played a role.

Also, it's important to acknowledge that the current investigation serves as an interim assessment of Plan Vélo I and II. Future studies, particularly upon full implementation, will provide valuable insights into traffic behavior and the overall effectiveness of the initiatives.



\section{Conclusions}

We analyzed the evolution of traffic behavior in Paris from 2010 to 2023 amidst significant network changes, utilizing the re-sampling method for NFDs based on empirical data. The interventions in Paris led to noticeable reductions in traffic flow in both zones studied, albeit to different extents.In Zone 1, higher congestion were observed, resulting primarily from reduced space for vehicles rather than an increase in traffic volume. 
Conversely, Zone 2 experienced lower congestion levels following the interventions. Zone 1, comprising of the central districts north of the Seine, represents a progressive transformation of the city, while Zone 2 embodies transformation endeavors more in line with the rest of the city.


In conclusion, the findings highlight that the interventions had a discernible effect. Traffic flow showed a significant decrease in both zones studied, varying depending on the specific zone analysed. It appears that in Zone 1, where more progressive measures were implemented, the rise in critical density and resulting higher congestion levels can be primarily attributed to a reduction in available space for vehicles, rather than an increase in traffic volume. 
Congestion has indeed increased, though not to the point of causing a complete traffic collapse. This suggests that demand responses likely shifted towards alternative modes of transportation. 

These effects can largely be attributed to policy interventions aimed at making the city more bicycle- and pedestrian-friendly, with Plan Vélo I and II playing a significant role, during which over 1,000 kilometres of new bicycle lanes were constructed. 

Paris' approach to smart mobility underscores the importance of closely monitoring network changes to inform effective traffic management decisions. Future research will continue to track network changes in Paris, particularly during the Paris Olympics in summer 2024. Additionally, we will focus on estimating changes on the demand side, such as trip rates and trip lengths, from the available data to enhance the perspective on this infrastructure-based transportation demand policy.

In closing, the revealed changes in the NFD, in particular in capacity and critical density underline that Paris underwent a substantial transformation from a car-oriented city to one that prioritizes cycling and public transport. Car travel is not banned but is disincentivized for many trips. As a result, it is not surprising that cycling has increased while car travel has decreased \cite{TrafficReductionParis}. However, the estimated NFDs indicate that Paris still experiences substantial congestion levels, as seen in the many observed traffic states beyond critical density in Figures \ref{fig:resampled_NFD_zone_1} and \ref{fig:resampled_NFD_zone_2}. 
Therefore, although the reallocation of urban space can be considered a successful transport demand management policy from a sustainability perspective, car traffic can be further improved by enhancing transport demand management policies, presumably through pricing-based strategies, e.g., based on the NFD \cite{zheng_dynamic_2012} or consumed mobility \cite{bliemer_novel_2024}.



\section{Acknowledgments}
Elena Natterer acknowledges funding from the project MINGA with grant number 45AOV1001K, by the German Federal Ministry of Transport and Digital Infrastructure. Allister Loder acknowledges support from the Bavarian State Ministry of Science and the Arts in the framework of the bidt Graduate Center for Postdocs. 

We thank Yamam Alayasreih for providing a first implementation of the re-sampling methodology and for fruitful discussions. 
ChatGPT 4.0 helped with spelling checks and summarising paragraphs. The authors remain responsible for all findings and opinions presented in the paper.

\section{Author contributions}

The authors confirm their contribution to the paper as follows. Study conception and design, analysis and interpretation of results, draft manuscript preparation: E. Natterer, A. Loder, K. Bogenberger; data collection and engineering: E. Natterer. All authors reviewed the results and approved the final version of the manuscript.


\bibliographystyle{trb}
\bibliography{references,bib_loder}

\begin{thebibliography}{38}
\providecommand{\natexlab}[1]{#1}

\bibitem[{par(2023)}]{paris_carbon_neutral}
\emph{City of Paris: Carbon Neutral by 2050 for a Fair, Inclusive and Resilient Transition | France}. https://unfccc.int/climate-action/un-global-climate-action-awards/climate-leaders/city-of-paris\#:~:text=By

\bibitem[{Buehler and Pucher(2022)}]{buehler}
Buehler, R. and J.~Pucher, Cycling through the COVID-19 pandemic to a more sustainable transport future: Evidence from case studies of 14 large bicycle-friendly cities in Europe and North America, 2022.

\bibitem[{pla(2023{\natexlab{a}})}]{plan_velo_1}
\emph{Plan Velo I}. https://www.paris.fr/pages/paris-a-velo-225, 2023{\natexlab{a}}.

\bibitem[{pla(2023{\natexlab{b}})}]{plan_velo_2}
\emph{Plan Velo II}. https://www.paris.fr/pages/un-nouveau-plan-velo-pour-une-ville-100-cyclable-19554, 2023{\natexlab{b}}.

\bibitem[{rem(2023)}]{remove_parking_spots}
\emph{Why Paris is eliminating 72 percent of its on-street parking spaces}. https://park4sump.eu/news-events/news/why-paris-eliminating-72-its-street-parking-spaces, 2023.

\bibitem[{O'Sullivan(2021)}]{paris_15_min_city}
O'Sullivan, L., Feargus;~Bliss, \emph{The 15-Minute City—No Cars Required—Is Urban Planning's New Utopia}. https://park4sump.eu/news-events/news/why-paris-eliminating-72-its-street-parking-spaces, 12 November 2020, retrieved 29 March 2021.

\bibitem[{{Paauwe, Maurits}(2021)}]{Paauwe_2021}
{Paauwe, Maurits}, \emph{{Assessing the Effect of Bike Lane Construction on Surrounding Property Values in Paris, France – A Quantitative Approach}}. Master's thesis, {rijksuniversiteit groningen}, 2021.

\bibitem[{Buehler and Pucher(2021)}]{cycling_book2021}
Buehler, R. and J.~Pucher (eds.) \emph{Cycling for Sustainable Cities}. The MIT Press, 2021.

\bibitem[{{Natterer, Elena and Loder, Allister and Bogenberger, Klaus}(2024)}]{TrafficReductionParis}
{Natterer, Elena and Loder, Allister and Bogenberger, Klaus}, Traffic Reduction and Decarbonization through Network Changes - Empirical Evidence from Paris. In \emph{{Presented at 96th Annual Meeting of the Transportation Research Board}}, 2024.

\bibitem[{Huang et~al.(2021)Huang, Sun, Li, and Axhausen}]{IBikeTrafficMFDShanghai}
Huang, Y., D.~Sun, A.~Li, and K.~W. Axhausen, Impact of bicycle traffic on the macroscopic fundamental diagram: some empirical findings in Shanghai. \emph{Transportmetrica A: Transport Science}, Vol. 426, 2021.

\bibitem[{Pucher et~al.(2009)Pucher, Dill, and Handy}]{ReviewInfrastructureToIncreaseBiking}
Pucher, J., J.~Dill, and S.~Handy, \emph{Infrastructure, programs, and policies to increase bicycling: An international review}. https://www.sciencedirect.com/science/article/pii/S0091743509004344, 2009.

\bibitem[{Haar(2023)}]{ImprovingCyclabilityInNetherlands}
Haar, F.~v., Improving the cyclability in The Netherlands: How infrastructural investments can the increase the bicycle share, 2023.

\bibitem[{Hull and O'~Holleran(2014)}]{BikeInfrastructureEncourageCycling}
Hull, A. and C.~O'~Holleran, \emph{Bicycle infrastructure: can good design encourage cycling?} https://www.tandfonline.com/doi/full/10.1080/21650020.2014.955210, 2014.

\bibitem[{van Goeverden et~al.(2015)van Goeverden, Sick~Nielsen, Harder, and van Nes}]{InterventionsInDutchCities}
van Goeverden, K., T.~Sick~Nielsen, H.~Harder, and R.~van Nes, Interventions in Bicycle Infrastructure, Lessons from Dutch and Danish Cases. \emph{Transportation Research Procedia}, 2015.

\bibitem[{Reggiani et~al.(2022)Reggiani, van Oijen, Hamedmoghadam, Daamen, Vu, and Hoogendoorn}]{UnderstandingBikeability}
Reggiani, G., T.~van Oijen, H.~Hamedmoghadam, W.~Daamen, H.~L. Vu, and S.~Hoogendoorn, \emph{Understanding bikeability: a methodology to assess urban networks}, 2022.

\bibitem[{Daganzo(2007)}]{Daganzo2007}
Daganzo, C.~F., Urban gridlock: {Macroscopic} modeling and mitigation approaches. \emph{Transportation Research Part B: Methodological}, Vol.~41, No.~1, 2007, pp. 49--62.

\bibitem[{Leclercq et~al.(2014)Leclercq, Chiabaut, and Trinquier}]{Leclercq2014}
Leclercq, L., N.~Chiabaut, and B.~Trinquier, Macroscopic fundamental diagrams: {A} cross-comparison of estimation methods. \emph{Transportation Research Part B: Methodological}, Vol.~62, 2014, pp. 1--12.

\bibitem[{Ambühl et~al.(2018)Ambühl, Loder, Bliemer, Menendez, and Axhausen}]{ambuhl_introducing_2018}
Ambühl, L., A.~Loder, M.~C. Bliemer, M.~Menendez, and K.~W. Axhausen, Introducing a {Re}-{Sampling} {Methodology} for the {Estimation} of {Empirical} {Macroscopic} {Fundamental} {Diagrams}. \emph{Transportation Research Record}, Vol. 2672, 2018, pp. 239--248.

\bibitem[{Mogridge(1997)}]{mogridge_self-defeating_1997}
Mogridge, M. J.~H., The self-defeating nature of urban road capacity policy. \emph{Transport Policy}, Vol.~4, 1997, pp. 5--23.

\bibitem[{Loder et~al.(2019{\natexlab{a}})Loder, Ambühl, Menendez, and Axhausen}]{loder_understanding_2019}
Loder, A., L.~Ambühl, M.~Menendez, and K.~W. Axhausen, Understanding traffic capacity of urban networks. \emph{Scientific Reports}, Vol.~9, No.~1, 2019{\natexlab{a}}, p. 16283.

\bibitem[{Castrillon and Laval(2018)}]{Castrillon2018}
Castrillon, F. and J.~Laval, Impact of buses on the macroscopic fundamental diagram of homogeneous arterial corridors. \emph{Transportmetrica B: Transport Dynamics}, Vol.~6, No.~4, 2018, pp. 286--301, publisher: Taylor \& Francis.

\bibitem[{Geroliminis et~al.(2014)Geroliminis, Zheng, and Ampountolas}]{Geroliminis2014}
Geroliminis, N., N.~Zheng, and K.~Ampountolas, A three-dimensional macroscopic fundamental diagram for mixed bi-modal urban networks. \emph{Transportation Research Part C: Emerging Technologies}, Vol.~42, 2014, pp. 168--181.

\bibitem[{Balzer and Leclercq(2024)}]{balzer_modal_2024}
Balzer, L. and L.~Leclercq, Modal dynamic equilibrium under different demand management schemes. \emph{Transportation}, Vol.~51, No.~2, 2024, pp. 529--566.

\bibitem[{Menelaou et~al.(2023)Menelaou, Timotheou, Kolios, and Panayiotou}]{menelaou_convexification_2023}
Menelaou, C., S.~Timotheou, P.~Kolios, and C.~G. Panayiotou, Convexification approaches for regional route guidance and demand management with generalized {MFDs}. \emph{Transportation Research Part C: Emerging Technologies}, Vol. 154, 2023, p. 104245.

\bibitem[{Geroliminis and Daganzo(2008)}]{Geroliminis2008}
Geroliminis, N. and C.~F. Daganzo, Existence of urban-scale macroscopic fundamental diagrams: {Some} experimental findings. \emph{Transportation Research Part B: Methodological}, Vol.~42, 2008, pp. 759--770.

\bibitem[{Daganzo and Geroliminis(2008)}]{Daganzo2008}
Daganzo, C.~F. and N.~Geroliminis, An analytical approximation for the macroscopic fundamental diagram of urban traffic. \emph{Transportation Research Part B: Methodological}, Vol.~42, 2008, pp. 771--781.

\bibitem[{Laval and Castrillón(2015)}]{laval_stochastic_2015}
Laval, J.~A. and F.~Castrillón, Stochastic approximations for the macroscopic fundamental diagram of urban networks. \emph{Transportation Research Part B: Methodological}, Vol.~81, No.~3, 2015, pp. 904--916, publisher: Elsevier B.V.

\bibitem[{Loder et~al.(2019{\natexlab{b}})Loder, Ambühl, Menendez, and Axhausen}]{Loder2019SciCap}
Loder, A., L.~Ambühl, M.~Menendez, and K.~W. Axhausen, Understanding traffic capacity of urban networks. \emph{Scientific Reports}, Vol.~9, No. 16283, 2019{\natexlab{b}}.

\bibitem[{Ortigosa et~al.(2017)Ortigosa, Gayah, and Menendez}]{ortigosa_analysis_2017}
Ortigosa, J., V.~V. Gayah, and M.~Menendez, Analysis of one-way and two-way street configurations on urban grid networks. \emph{Transportmetrica B: Transport Dynamics}, Vol. in press, 2017, pp. 1--21, publisher: Taylor \& Francis.

\bibitem[{Dantsuji et~al.(2021)Dantsuji, Fukuda, and Zheng}]{Dantsuji2021}
Dantsuji, T., D.~Fukuda, and N.~Zheng, Simulation-based joint optimization framework for congestion mitigation in multimodal urban network: a macroscopic approach. \emph{Transportation}, Vol.~48, No.~2, 2021, pp. 673--697, publisher: Springer US ISBN: 0123456789.

\bibitem[{Loder et~al.(2022)Loder, Bliemer, and Axhausen}]{loder_optimal_2022}
Loder, A., M.~C. Bliemer, and K.~W. Axhausen, Optimal pricing and investment in a multi-modal city — {Introducing} a macroscopic network design problem based on the {MFD}. \emph{Transportation Research Part A: Policy and Practice}, Vol. 156, No. January, 2022, pp. 113--132.

\bibitem[{Ambühl et~al.(2021)Ambühl, Loder, Leclercq, and Menendez}]{Ambuhl2021}
Ambühl, L., A.~Loder, L.~Leclercq, and M.~Menendez, Disentangling the city traffic rhythms: {A} longitudinal analysis of {MFD} patterns over a year. \emph{Transportation Research Part C: Emerging Technologies}, Vol. 126, 2021, p. 103065, publisher: Pergamon.

\bibitem[{Ambühl et~al.(2017)Ambühl, Loder, Menendez, and Axhausen}]{Ambuhl2017}
Ambühl, L., A.~Loder, M.~Menendez, and K.~W. Axhausen, Empirical macroscopic fundamental diagrams: {New} insights from loop detector and floating car data. In \emph{96th {Annual} {Meeting} of the {Transportation} {Research} {Board}}, Washington D.C., 2017.

\bibitem[{Par(2024)}]{ParisOpenData_Cycleways}
\emph{Official cycle paths Paris}. https://opendata.paris.fr/explore/dataset/reseau-cyclable/information/, 2024.

\bibitem[{Kara(2024)}]{kara2024}
Kara, E., \emph{Analyse der Fahrradnetzwerkänderungen in Paris, Frankreich}. Bachelor's thesis, https://www.mos.ed.tum.de/fileadmin/w00ccp/vt/theses/pdf/BA/2024/BA-Kara-Aushang-Jun24.pdf, 2024.

\bibitem[{Pla(2024)}]{PlanVeloII_ParisOpenData}
\emph{Official Plan Vélo II}. https://opendata.paris.fr/explore/dataset/plan-velo-2026/table/, 2024.

\bibitem[{Zheng et~al.(2012)Zheng, Waraich, Axhausen, and Geroliminis}]{zheng_dynamic_2012}
Zheng, N., R.~A. Waraich, K.~W. Axhausen, and N.~Geroliminis, A dynamic cordon pricing scheme combining the {Macroscopic} {Fundamental} {Diagram} and an agent-based traffic model. \emph{Transportation Research Part A: Policy and Practice}, Vol.~46, No.~8, 2012, pp. 1291--1303.

\bibitem[{Bliemer et~al.(2024)Bliemer, Loder, and Zheng}]{bliemer_novel_2024}
Bliemer, M.~C., A.~Loder, and Z.~Zheng, A novel mobility consumption theory for road user charging. \emph{Transportation Research Part B: Methodological}, 2024, p. 102998.

\end{thebibliography}

\end{document}